\documentclass[seceq]{ptptex}

\usepackage{graphicx}
\input epsf
\usepackage{wrapft}

\newcommand{\A}{{\mathcal{A}}}

\newcommand{\lsim}   {\mathrel{\mathop{\kern 0pt \rlap
  {\raise.2ex\hbox{$<$}}}
  \lower.9ex\hbox{\kern-.190em $\sim$}}}
\newcommand{\gsim}   {\mathrel{\mathop{\kern 0pt \rlap
  {\raise.2ex\hbox{$>$}}}
  \lower.9ex\hbox{\kern-.190em $\sim$}}}

\title{
On the electromagnetic nature of dark energy and the origin of
cosmic magnetic fields
}

\author{
Jose Beltr\'an Jim\'enez$^{1,2}$\footnote{E-mail: Jose.Beltran@unige.ch} 
and Antonio L. Maroto$^{2,}$\footnote{E-mail: maroto@fis.ucm.es}
}

\inst{
$^1$Institute de Physique Th\'eorique, Universit\'e de Gen\`eve, 
24 quai E. Ansermet, 1211 Gen\`eve 4, Switzerland\\
$^2$Departamento de F\'isica Te\'orica, Universidad Complutense 
de Madrid, 28040, Madrid, Spain
}

\abst{
In this work we consider quantum electromagnetic fields in an 
expanding universe.
We start by reviewing the difficulties found when trying to impose
the Lorenz condition in a time-dependent geometry.
Motivated by this fact, we explore the possibility of 
extending the electromagnetic theory by allowing the 
scalar state which is usually eliminated by means
of the Lorenz condition to propagate,  
preserving at the same time 
the dynamics of ordinary transverse photons. We show that  
the new state cannot be generated by charged currents, but it
breaks conformal invariance and can be excited gravitationally. 
In fact, primordial 
quantum fluctuations produced during inflation can give 
rise to super-Hubble temporal electromagnetic 
modes whose energy density behaves as a cosmological constant.
The value of the effective cosmological constant is shown to 
agree with observations provided inflation took place 
at the electroweak scale. The theory is compatible with all the 
local gravity tests and is free from classical or quantum 
instabilities. Thus we see that, not only the true nature of dark energy 
can be established without resorting to new physics, but also 
the value of the cosmological constant finds a natural explanation 
in the context of standard inflationary cosmology. 
On sub-Hubble scales, the new state generates an effective
 charge density which,  due to the high 
electric conductivity of the cosmic plasma after inflation, 
gives rise to both vorticity and magnetic fields. 
Present upper limits on vorticity coming from CMB anisotropies 
are translated into lower limits on the present value of cosmic 
magnetic fields. We find that magnetic fields 
$B_{\lambda}> 10^{-12}$ G can be typically generated 
with coherence lengths ranging from sub-galactic scales up to the present 
Hubble radius. Those fields could act as seeds for a galactic dynamo or even 
account for observations just by collapse and differential rotation of the 
protogalactic cloud.}

\begin{document}
\maketitle

\section{Introduction}

Maxwell's electromagnetism and its quantum counterpart, Quantum
Electrodynamics (QED), provide a very accurate description 
of electromagnetic propagation and interactions in a huge range
of scales, from the extremely small distances probed by high-energy 
colliders and cosmic rays, up to the coherence lengths around 1.3 A.U. of 
the magnetic fields dragged by the solar wind \cite{limit}. 
However, the behaviour of electromagnetic fields with
wavelengths larger than the solar system radius is still far from clear. Indeed,
not only we do not have experimental access to such a low energy 
regime, but more importantly, the standard theory does not
seem to be able to explain the 
origin of the cosmic magnetic fields found in galaxies, clusters\cite{galactic} 
and, 
very recently\cite{extragalactic}, also in the voids.   

A similar situation is found for gravity, the other long-range
interaction in nature. Although General Relativity (GR) provides a 
good description of gravity from sub-millimeter scales up to 
solar system distances, several problems on larger scales 
have led to postulate the existence of exotic components in 
the universe, i.e. dark matter and dark energy, or
even to consider modifications of GR on cosmological scales.

The above discussion suggests that a more careful analysis of the behaviour
of electromagnetic fields in cosmological contexts is needed. 
A fundamental aspect of this problem which has not received
much attention is the issue of  quantization of  
 gauge theories in time-dependent geometries.
As is well-known, due to the redundancy in the 
description of gauge fields, all the quantization approaches 
 restrict in some way the possible physical states of the theory.  
Thus, for example, most of the studies performed in  expanding backgrounds  
have been based in the use of the Coulomb gauge fixing condition. 
In flat space-time, Coulomb quantization is 
equivalent to other methods such as covariant (Gupta-Bleuler) or
BRST quantization. However, in a curved space-time this equivalence
has not been proved. As a matter of fact, Gupta-Bleuler formalism
has been shown to be ill-defined in a time-dependent background
\cite{Parker,EM2}, and also BRST method has been shown
to exhibit similar pathologies in certain space-time geometries \cite{ghosts}.
These difficulties are rooted in the impossibility of uniquely decomposing
the fields in their positive and negative frequency modes in a curved
space-time. This fact, in turn, prevents a unique definition of the space 
of physical states in those formalisms.

In this work, we explore an extended theory of electromagnetism
which avoids the above mentioned problems by allowing the propagation
of the state which is usually eliminated by means of the subsidiary condition. 
The price to pay is the modification of Maxwell's 
equations. However, this modification does not affect
the dynamics of ordinary transverse photons and, what is more
remarkable, it introduces an
effective current in the equations which, unlike ordinary equations, 
allow the generation of cosmic magnetic fields 
all the way from sub-galactic scales up to the present Hubble radius
\cite{magnetic}.
On the other hand, on super-Hubble scales, it can be seen that the new
state contributes as an effective cosmological constant \cite{EM1}. 
Interestingly, 
because of the breaking of conformal triviality in the modified
theory, the new state can be excited from quantum fluctuations
during inflation. Thus, in a completely analogous way to the generation
of metric and density perturbations, in the extended theory, an 
almost scale invariant spectrum of electromagnetic perturbations is
also produced.
The correct amplitude required to explain the present
phase of accelerated expansion of the universe and the 
observed cosmic magnetic fields can be obtained in a natural way provided 
inflation took place at the electroweak scale.    

\section{Covariant quantization in flat space-time}

Let us start by briefly reviewing the standard covariant 
quantization method in Minkowski space-time \cite{Itzykson}
since this will be useful in the rest of the work.
The starting point is the modified electromagnetic action:
\begin{eqnarray}
S=\int d^4x \left(-\frac{1}{4}F_{\mu\nu}F^{\mu\nu}+\frac{\xi}{2}
(\partial_\mu A^\mu)^2+ A_\mu J^\mu\right).
 \label{actionGB}
\end{eqnarray}
Because of the presence of the gauge breaking $\xi$-term, 
this action is no longer invariant under arbitrary 
gauge transformations, but
only under residual ones given by: 
$A_\mu\rightarrow A_\mu+\partial_\mu \theta$, provided $\Box
\theta=0$.
The equations of motion obtained from this action now read:
\begin{eqnarray}
\partial_\nu F^{\mu\nu}+\xi\partial^\mu(\partial_\nu
A^\nu)=J^\mu.
\label{fieldeq}
\end{eqnarray}
In order to recover ordinary Maxwell's equation, the Lorenz condition 
$\partial_\mu A^\mu=0$ must
be imposed so that the $\xi$-term disappears. At the classical
level this can be achieved by means of appropriate boundary
conditions on the field. Indeed, taking the four-divergence of the
above equation, we find:
\begin{eqnarray}
\Box(\partial_\nu A^\nu)=0
\end{eqnarray}
where we have made use of current conservation. This means that
the field  $\partial_\nu A^\nu$ evolves as a free scalar field, so
that if it vanishes for large $\vert t \vert$, it will vanish at 
all times. At the quantum level, the Lorenz condition cannot be
imposed as an operator identity, but only in the weak sense
$\partial_\nu A^{\nu \,(+)}\vert \phi\rangle=0$, where $(+)$
denotes the positive frequency part of the operator and $\vert
\phi\rangle$ is a physical state. This condition is
equivalent to imposing $[{\bf a}_0(\vec k) +{\bf a}_\parallel(\vec
k)] |\phi\rangle=0$, with ${\bf a}_0$ and ${\bf a}_\parallel$ the
annihilation operators corresponding to temporal and longitudinal
electromagnetic states. Thus, in the covariant formalism, the
physical states contain the same number of temporal and
longitudinal photons, so that their energy densities, having
opposite signs, cancel each other. Therefore, only the 
transverse photons contribute to the energy density.

\section{Covariant quantization in an expanding universe}

Let us  consider the curved
space-time version of action (\ref{actionGB}):
\begin{eqnarray}
S=\int d^4x
\sqrt{g}\left[-\frac{1}{4}F_{\mu\nu}F^{\mu\nu}+\frac{\xi}{2}
(\nabla_\mu A^\mu)^2+ A_\mu J^\mu\right].
 \label{actionF}
\end{eqnarray}
Now the modified Maxwell's equations read:
\begin{eqnarray}
\nabla_\nu F^{\mu\nu}+\xi\nabla^\mu(\nabla_\nu A^\nu)=J^\mu
\label{EMeqexp}
\end{eqnarray}
and taking again the four divergence, we get:
\begin{eqnarray}
\Box(\nabla_\nu A^\nu)=0.\label{minimal}
\end{eqnarray}
We see that once again  $\nabla_\nu A^\nu$  behaves as a scalar
field which is decoupled from the conserved electromagnetic
currents, but it is non-conformally coupled to gravity. This means
that, unlike the flat space-time case,  this field can be excited
from quantum vacuum fluctuations by the expanding background and  
this poses the question of the validity of the
Lorenz condition at all times. Thus, for example, 
let us consider as a toy example the quantization in 
an expanding background interpolating between two
asymptotically flat regions: $ds^2=a(\eta)^2(d\eta^2-d\vec x^2)$ with
$a(\eta)=2+\tanh(\eta/\eta_0)$ where $\eta_0$ is constant.

\begin{figure}
\begin{center}
  {\epsfxsize=7.5cm\epsfbox{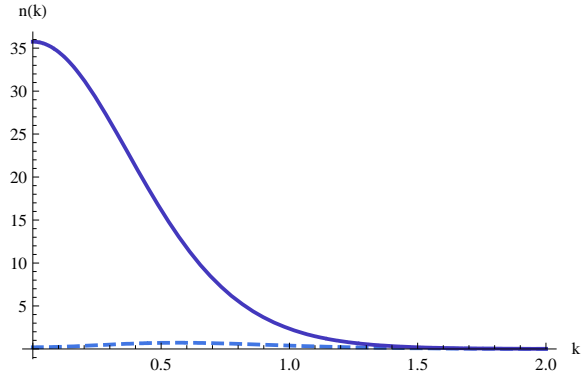}}
  \caption{Occupation numbers for temporal (continuous
line) and longitudinal (dashed line) photons 
in the $out$ region vs.
$k$ in $\eta_0^{-1}$ units.}
\end{center}
\end{figure}

 If we prepare our
system  in an initial  state $\vert \phi\rangle$ belonging
to the physical Hilbert space, for example with 
$n_0^{out}(k)= n_\parallel^{out}(k)=0$, 
i.e. satisfying $\partial_\nu \A^{\nu \,(+)}_{in}\vert \phi\rangle=0$
in the initial flat region.
Because of the expansion
of the universe, the positive frequency modes in the $in$
region with a given temporal or longitudinal polarization 
will become a linear superposition of positive and
negative frequency modes in the $out$ region and
with different polarizations \cite{EM2}.
Thus, the system will end up in a final state which no longer
satisfies the weak Lorenz condition i.e. in the {\it out} region
$n_0^{out}(k)\neq n_\parallel^{out}(k)$ or in other words, 
$\partial_\nu \A^{\nu \,(+)}_{out}\vert \phi\rangle\neq 0$
as shown in Fig. 1.

A similar problem with the subsidiary conditions has been recently found 
in BRST quantization in Rindler space-time \cite{ghosts}. 
Thus, in this work   we will explore
the possibility of quantization  in an expanding
universe without imposing the Lorenz condition.

\section{Extended electromagnetism without the Lorenz condition}

Let us then explore the possibility that the fundamental 
theory of electromagentism is  given 
 by the modified action (\ref{actionF}) where we allow
the $\nabla_\mu A^\mu$ field to propagate. 
Since we are not imposing the Lorenz condition, in principle,
 important viability problems for the theory could arise, 
namely: modification of classical Maxwell's equations, 
new unobserved photon polarizations, negative norm (energy) states or 
 conflicts with QED phenomenology. 
However, as we will show in the following, none of these 
problems is actually present. 

Having removed one constraint, 
the theory  contains one additional degree of freedom. Thus, 
the general solution for the modified equations (\ref{EMeqexp})
can be written as:
\begin{eqnarray}
\A_\mu=\A_\mu^{(1)}+\A_\mu^{ (2)}+\A_\mu^{(s)}+\partial_\mu \theta
\end{eqnarray}
where $\A_\mu^{(i)}$ with $i=1,2$ are the two transverse modes of
the massless photon, $\A_\mu^{(s)}$ is the new scalar state, which
is the mode that would have been eliminated if we had imposed the
Lorenz condition and, finally, $\partial_\mu \theta$ is a purely
residual gauge mode, which can be eliminated by means of a
residual gauge transformation in the asymptotically free regions, 
in a completely analogous way to the elimination of the $A_0$
component in the Coulomb or Lorenz gauge quantization.  The fact that
Maxwell's electromagnetism could contain an additional scalar
mode decoupled from electromagnetic currents, but with 
 non-vanishing  gravitational interactions, was already noticed 
in a different context in \cite{Deser}. 

In order to quantize the free theory, we perform the mode
expansion of the  field with the corresponding creation and
annihilation operators for the {\it three} physical states:
\begin{eqnarray}
\A_{\mu}=\int d^3\vec{k}\ \sum_{\lambda=1, 2,s}\left[{\bf
a}_\lambda(k)\A_{\mu k}^{(\lambda)} +{\bf
a}_\lambda^\dagger(k)\overline{\A_{\mu k}^{(\lambda)}}\, \right]
\end{eqnarray}
where the modes are required to be orthonormal with respect to the
appropriate scalar product. Notice that
the three modes can be chosen to have positive normalization, 
and therefore:
\begin{eqnarray}
\left[{\bf a}_\lambda(\vec{k}),{\bf a}_{\lambda'}^\dagger(\vec{k'})\right]
=\delta_{\lambda\lambda'}\delta^{(3)}(\vec{k}-\vec{k'}),\;\;\;
\lambda,\lambda'=1,2,s
\end{eqnarray}
We see that the sign of the commutators is positive for the
three physical states, i.e. the negative norm state can be
eliminated in the free theory.

The evolution of the new mode is given by (\ref{minimal}), so that  
 on super-Hubble scales ($\vert k\eta\vert \ll 1$),
$\vert\nabla_\mu\A^{(s)\mu}_k\vert= const.$ which, as shown in
\cite{EM1}, implies that the field contributes as a cosmological
constant in (\ref{actionF}). Indeed, the energy-momentum tensor 
derived from (\ref{actionF}) reads in that limit:
\begin{eqnarray}
T_{\mu\nu}=\frac{\xi}{2} g_{\mu\nu}(\nabla_\alpha A^\alpha)^2
\end{eqnarray}
which is the energy-momentum tensor of a cosmological constant.  
Notice that,  as seen in (\ref{minimal}), the new scalar mode is a 
massless free field and it is possible to calculate the corresponding 
power spectrum  generated during
inflation, 
$P_{\nabla A}(k)=4\pi k^3\vert\nabla_\mu\A^{(s)\mu}_k\vert^2 $. In the
super-Hubble limit, we get  in a 
 quasi-de Sitter inflationary phase characterized by a slow-roll
parameter $\epsilon$:
\begin{eqnarray}
P_{\nabla A}(k)=\frac{9H_{k_0}^4}{16\pi^2}
\left(\frac{k}{k_0}\right)^{-4\epsilon}
\label{PE}
\end{eqnarray}
where $H_{k_0}$ is the Hubble parameter when the 
$k_0$ mode left the horizon and we have chosen 
the normalization so that $\xi=1/3$ (see \cite{EM1}). Notice that this
result implies that $\rho_A\sim
\frac{9H_{k_0}^4}{64\pi^2\epsilon}\left(\frac{H_0}{k_c}\right)^{4\epsilon}$. 
Taking  the infrared cutoff $k_c$ 
as the comoving Hubble radius at the beginning of inflation
(see \cite{Tanaka} and references therein for problems with infrared
divergences during inflation), the measured value of
the cosmological constant then typically requires 
$H_{k0}\simeq 2 \times 10^{-6}$ eV,
which corresponds to an inflationary scale  $M_I\sim 100$ GeV.
Thus we see that the cosmological constant scale can be naturally
explained in terms of physics at the electroweak scale.
This is one of the most relevant aspects of the present model
in which, unlike existing dark energy theories based on scalar fields, 
dark energy can be generated without including any potential term
or dimensional constant.

As shown above, the field amplitude
remains frozen on super-Hubble scales, so that no modification
of Maxwell's equation is generated on those scales, however as the
amplitude starts decaying once the
mode enters the horizon in the radiation or matter eras, 
 the $\xi$ term in (\ref{EMeqexp}) generates an effective
current which can produce magnetic fields on cosmological
scales, as we will show below.

Notice that in Minkowski space-time, the  theory (\ref{actionF})
is completely equivalent to standard QED. This is so because, although
non-gauge invariant, the corresponding effective action is equivalent to the 
standard BRST invariant effective action of QED \cite{EM2}.

To summarize, none of the above mentioned consistency problems for the theory
in (\ref{actionF}) arise, thus: the new state can only be generated 
gravitationally and evades laboratory detection; the new state has 
positive norm (energy); the effective action is completely equivalent 
to standard QED in the flat space-time limit and although ordinary 
Maxwell's equations are modified on small scales,
the only effect of the new term is the generation of cosmic magnetic fields.

On the other hand, despite the fact that the background evolution in the present case
is the same as
in $\Lambda$CDM, the evolution of metric perturbations could
be different. We have calculated the evolution of metric,
matter density and
electromagnetic perturbations \cite{EM3}. The propagation speeds
of scalar, vector and tensor perturbations are found
to be real and equal to the speed of light, so that the theory is
classically stable.
On the other hand, it is
possible to see that all the parametrized post-Newtonian (PPN) parameters
\cite{Will}
agree with those of General Relativity,  i.e. the theory is compatible
with all the local gravity constraints for any value
of the homogeneous background electromagnetic field \cite{EM1,viability}.

Concerning the evolution of
scalar perturbations, 
we find
that  the only relevant deviations with respect to $\Lambda$CDM
appear on large scales ($k\sim H_0$) and that
they depend on the primordial
spectrum of electromagnetic fluctuations. However,
the effects on the CMB temperature and matter power spectra 
are compatible with observations except for very large primordial
fluctuations \cite{EM3}.

\section{Generation of cosmic magnetic fields}

It is interesting to note that the $\xi$-term 
can be seen, at the equations of motion
level, as a conserved current acting as a source of the usual
Maxwell field. To see this, we can write
$-\xi\nabla^\mu(\nabla_\nu A^\nu)\equiv J_{\nabla\cdot A}^\mu$
which, according to (\ref{minimal}), satisfies the conservation
equation $\nabla_\mu J_{\nabla\cdot A}^\mu=0$ and we can express
(\ref{EMeqexp}) as:
\begin{eqnarray}
\nabla_\nu F^{\mu\nu}=J^\mu_T
\end{eqnarray}
with $J^\mu_T=J^\mu+J^\mu_{\nabla\cdot A}$ and $\nabla_\mu
J^\mu_T=0$. Physically, this means that, while the new scalar mode
can only be excited gravitationally, once it is
produced it will generally behave as a source of electromagnetic
fields. Therefore,  the modified theory is described by 
ordinary Maxwell equations with an additional "external" current.
For an observer with four-velocity
$u^\mu$ moving with the cosmic plasma, 
it is possible to  decompose the Faraday tensor in its 
electric and magnetic parts  
as: $F_{\mu\nu}=2E_{[\mu}u_{\nu ]}+\frac{\epsilon_{\mu\nu\rho\sigma}}
{\sqrt{g}}B^\rho u^\sigma$, where $E^\mu=F^{\mu\nu} u_\nu$ and
 $B^\mu=\epsilon^{\mu\nu\rho\sigma} /(2\sqrt{g})F_{\rho\sigma}u_\nu$.
Due to the infinite conductivity of the plasma, 
Ohm's law $J^\mu-u^\mu u_\nu J^\nu=\sigma F^{\mu\nu} u_\nu$
implies  $E^\mu=0$.  
Therefore, in that case the only contribution  would come
from  the magnetic part.  
Thus, from Maxwell's equations, we get:
\begin{eqnarray}
F^{\mu\nu}_{\;\;\;\; ;\nu}u_{\mu}=
\frac{\epsilon^{\mu\nu\rho\sigma}}
{\sqrt{g}}B_\rho u_{\sigma \,;\nu}u_\mu=J^{\mu}_{\nabla\cdot A}u_{\mu}
\end{eqnarray}
that for  comoving observers in a FLRW metric imply
(see also \cite{FC}):
\begin{eqnarray}
\vec \omega\cdot \vec B=\rho_g^0
\label{mag}
\end{eqnarray}
where $\vec v=d\vec x/d\eta$ is the conformal
time fluid velocity, $\vec\omega=\vec \nabla\times \vec v$ is 
the fluid 
vorticity, $\rho_g^0$ is the effective charge density today   
and the $\vec B$ components scale as $B_i\propto 1/a$ as can be
easily obtained from $\epsilon^{\mu\nu\rho\sigma}F_{\rho\sigma;\nu}=0$
to the lowest order in $v$. Thus,  the 
presence of the non-vanishing cosmic effective charge density necessarily 
creates both magnetic field and vorticity, in fact, 
we find that vorticity grows as $\vert\vec\omega\vert
\propto a$, from radiation era until present.

Notice that $\nabla_\nu A^\nu$ is constant on super-Hubble scales
 and starts
decaying as $1/a$ once the mode reenters the Hubble radius. Thus, today, 
a mode $k$ will have been suppressed by a 
factor  $a_{in}(k)$ (we are assuming that the 
scale factor today is $a_0=1$). This factor will be given by: 
$a_{in}(k)=\Omega_M H_0^2/k^2$ for 
modes entering the Hubble radius in the matter era, i.e. for 
$k<k_{eq}$ with $k_{eq}\simeq (14\, \mbox{Mpc})^{-1}\Omega_Mh^2$ 
the value of the mode which entered
at matter-radiation equality. For $k>k_{eq}$ we have 
$a_{in}(k)=\sqrt{2\Omega_M}(1+z_{eq})^{-1/2}H_0/k$.
It is then possible
to compute from (\ref{PE}) 
the corresponding power spectrum for the effective 
electric charge
density today $\rho_g^0=J_{\nabla\cdot A}^0=-\xi\partial_0(\nabla_\nu A^\nu)$.
Thus from:
\begin{eqnarray} 
\langle\rho(\vec k)\rho^*(\vec h)\rangle=
(2\pi)^3\delta(\vec k-\vec h)\rho^2(k)
\end{eqnarray}
we define 
$P_\rho(k)=\frac{k^3}{2\pi^2}\rho^2(k)$, which is given by:
\begin{eqnarray}
P_\rho(k)=\left\{
\begin{array}{cc}
0, & k<H_0\\
& \\
\frac{\Omega_M^2H_0^2 H_{k0}^4}{16\pi^2}
\left(\frac{k}{k_0}\right)^{-4\epsilon-2},& H_0<k<k_{eq}\\
&\\
\frac{2\Omega_M H_0^2 H_{k0}^4}{16\pi^2(1+z_{eq})}
\left(\frac{k}{k_0}\right)^{-4\epsilon},&  k>k_{eq}.
\end{array}\right. 
\label{chPE}
\end{eqnarray}
Therefore the corresponding charge variance will read:
 $\langle\rho^2\rangle=\int \frac{dk}{k}P_\rho(k)$.
Notice that for modes entering the Hubble radius in the
radiation era, the power spectrum is nearly scale
invariant. Also, due to the constancy of $\nabla_\nu A^\nu$ on super-Hubble 
scales, 
the effective charge density power spectrum is negligible on such scales, so that
we do not expect magnetic field nor vorticity generation on those
scales. Notice that, on sub-Hubble scales, the effective charge density
generates longitudinal electric fields 
whose present amplitude would be precisely $E_L\simeq \nabla_\nu A^\nu$.

\begin{figure}[ht!]
\begin{center}
{\epsfxsize=6.8cm\epsfbox{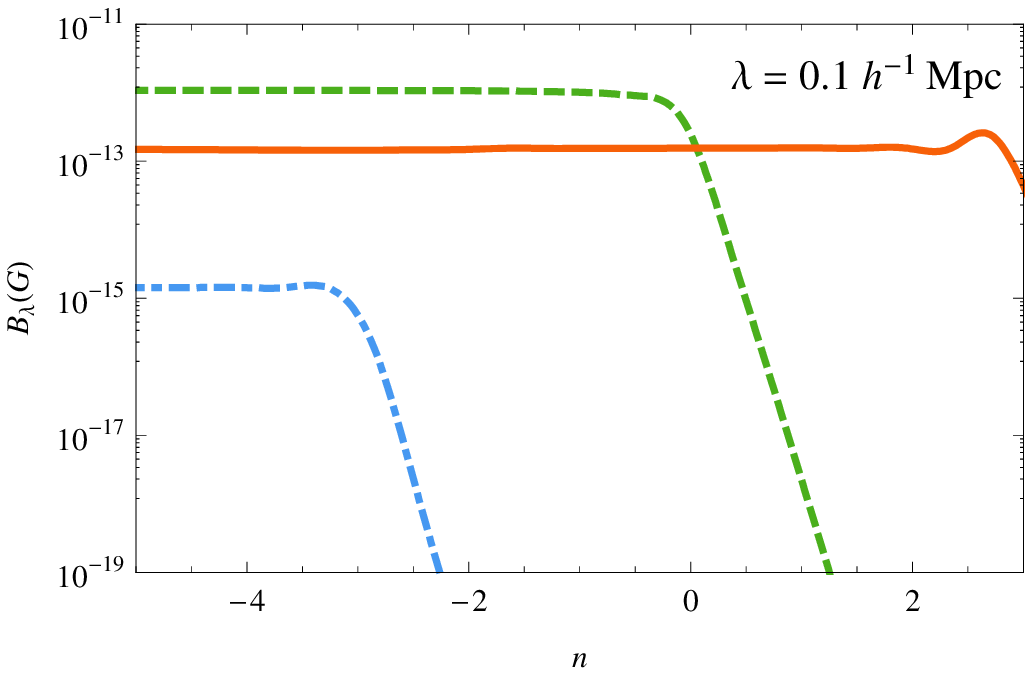}}
{\epsfxsize=6.8cm\epsfbox{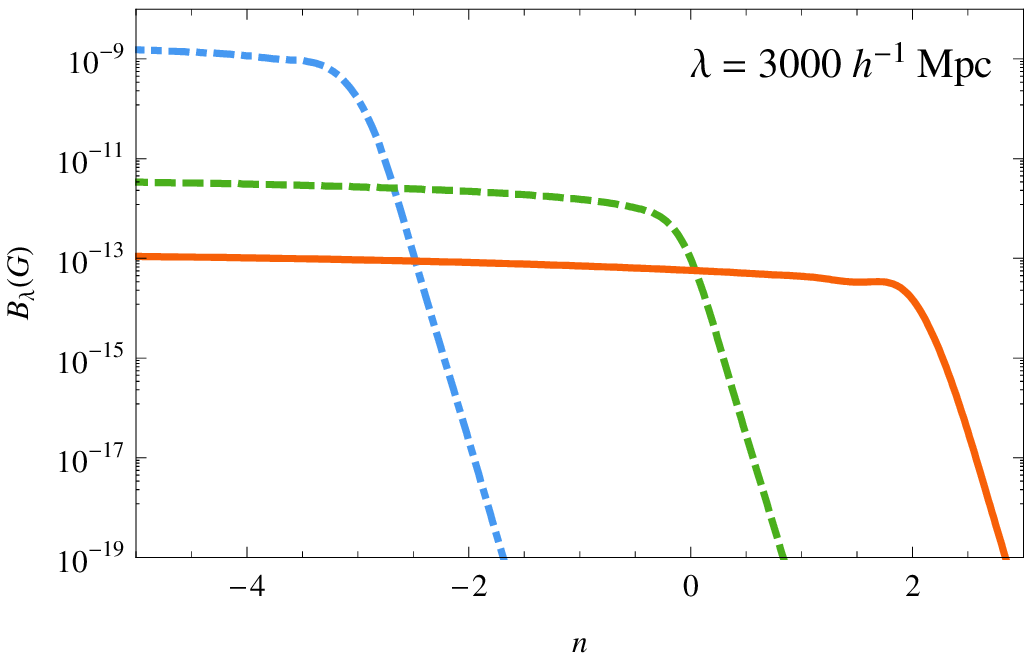}} 
\vspace*{-0.3cm}
\caption{ Lower limits on the magnetic
fields generated  on galactic scales (left panel)
and Hubble horizon scales (right panel) in terms
of the magnetic spectral index $n$ for different values
of the vorticity spectral index $m$. Dot-dashed blue for $m=0$, dashed 
green
for $m\simeq -3$ and full red for $m\simeq -5$.}
\end{center}
\end{figure} 
 Using (\ref{mag}), it is possible to translate the existing
upper limits on vorticity coming from CMB anisotropies \cite{FC} into  
{\it lower} limits on the amplitude of the magnetic fields 
generated by this mechanism. We will consider for simplicity
magnetic field and vorticity as gaussian stochastic variables
such that:
\begin{eqnarray}
\langle B_i(\vec k)B_j^*(\vec h)\rangle&=&
\frac{(2\pi)^3}{2}P_{ij}\delta(\vec k-\vec h)B^2(k)\nonumber \\ 
\langle\omega_i(\vec k)\omega_j^*(\vec h)\rangle&=&
\frac{(2\pi)^3}{2}P_{ij}\delta(\vec k-\vec h)\omega^2(k)
\end{eqnarray}
with $B^2(k)=B k^n$, $\omega^2(k)=\Omega k^m$
and where $P_{ij}=\delta_{ij}-\hat k_i\hat k_j$ is 
introduced because of the 
transversality properties of $B_i$ and $\omega_i$.
The spectral indices $n$ and $m$ are in principle arbitrary. 
In Fig. 2 we show  the 
lower limits on the magnetic fields
generated by this mechanism on  scales $\lambda=0.1 h^{-1}$ Mpc,
and  $\lambda=3000 h^{-1}$ Mpc, also for inflation at the
electroweak scale.     
 We see that fields can be generated with sufficiently 
large amplitudes in order to seed a galactic
dynamo or even to account for observations  just by collapse
and differential rotation of the protogalactic cloud \cite{magnetic}.
Moreover, they could be also compatible with recent extra-galactic 
observations\cite{extragalactic}.

\section{Discussion}
In this work, we have shown how a minimal extension of electromagnetism
which does not require the introduction of new fields, dimensional parameters
or potential terms  could provide a simple explanation for
the tiny value of the cosmological constant and, at the same time, a
mechanism for the generation of magnetic fields on cosmological scales.

Additional aspects of the theory, such as the modification of the 
magnetohydrodynamical evolution, should be considered in more detail
in order to make more precise predictions of the magnetic field amplitudes
for different astrophysical objects. Also a complete study 
of the quantum theory including interactions would help understanding
the physical properties of the new electromagnetic mode.
In any case, the results presented in this work provide a
clear example of  the potential cosmological 
implications which could appear in the  
interplay between gravity and quantum gauge theories.

\vspace{0.2cm}

{\em Acknowledgments:}
 This work has been  supported by
MICINN (Spain) project numbers
FIS 2008-01323 and FPA
2008-00592, CAM/UCM 910309, 
MEC grant BES-2006-12059 and MICINN Consolider-Ingenio 
MULTIDARK CSD2009-00064.

\end{document}